\documentstyle[12pt,aasms4]{article}
\begin{document}
\font\ninerm=cmr9
\lefthead{Struck \& Smith}
\righthead{Simple Models of Turbulent Disks}

%\begin{document}
\title{Simple Models for Turbulent Self-Regulation in Galaxy Disks}

\author{Curtis Struck, and Daniel C. Smith}
\affil{Dept. of Physics \& Astronomy, Iowa State University, Ames, IA 50011}
\authoremail{curt@iastate.edu & dcs@iastate.edu}

\begin{abstract}

Supernova explosions, and winds and energetic photon fluxes from young
star clusters drive outflows and supersonic turbulence in the
interstellar medium in galaxy disks, and provide broad spectrum
heating which generates a wide range of thermal phases in the gas.
Star formation, the source of the energy inputs, is itself regulated
by heating and phase exchanges in the gas.  However,
thermohydrodynamic self-regulation cannot be a strictly local process
in the interstellar gas, since galaxy disks also have a nearly
universal structure on large scales.

We propose that turbulent heating, wave pressure and gas exchanges
between different regions of disks play a dominant role in determining
the preferred, quasi-equilibrium, self-similar states of gas disks on
large-scales.  In this paper we present simple families of analytic,
thermohydrodynamic models for these global states, which include terms
for turbulent pressure and Reynolds stresses.  In these model disks
star formation rates, phase balances, and hydrodynamic forces are all
tightly coupled and balanced.  The models have stratified radial
flows, with the cold gas slowly flowing inward in the midplane of the
disk, and with the warm$/$hot phases that surround the midplane
flowing outward.

The models suggest a number of results that are in accord with
observation, as well as some novel predictions, including the
following.  1) The large-scale gas density and thermal phase
distributions in galaxy disks can be explained as the result of
turbulent heating and spatial couplings.  2) The turbulent pressures
and stresses that drive radial outflows in the warm gas above and
below the disk midplane also allow a reduced circular velocity
there. This effect was observed by Swaters, Sancisi \& van der Hulst
in NGC 891, a particularly turbulent edge-on disk. The models predict
that the effect should be universal in such disks. 3) Since
dissipative processes generally depend on the square of the gas
density, the heating and cooling balance in these models requires a
star formation rate like that of the Schmidt Law.  Conversely, they
suggest that the Schmidt Law is the natural result of global
thermohydrodynamical balance, and may not obtain in disks far from
equilibrium. 

\end{abstract}


\section{Introduction: Interstellar Gas in Galaxy Disks}
\subsection{Phase Structure and Turbulence}

The recognition that supernovae, massive star winds, and other
impulsive energy inputs are important heat sources for the
interstellar medium (ISM) in galaxies, and that they generate the warm
and hot phases, was a turning point for the theory of the ISM in
galaxies (e.g., McKee \& Ostriker 1977).  Studies of giant expanding
gas shells in our galaxy (e.g., Heiles 1984), and other nearby disk
galaxies showed that the energies required to produce these structures
are much greater than those produced by a single supernova or the
winds of its stellar progenitor (see reviews of Brinks 1990, van der
Hulst 1996).  However, young clusters of hot stars and their multiple
supernova do produce sufficient energy to make these supershells
(e.g., Mac Low, McCray \& Norman 1989, Norman \& Ikeuchi 1989,
Tenorio-Tagle, Rozyczka, \& Bodenheimer 1990, and Tomisaka 1992).
With this realization the theory of superbubbles, breakout, chimneys
and large-scale outflows developed quickly, and joined the older
theory of galactic fountains (see Shapiro \& Field 1976, Bregman 1980,
Cox 1981, Shapiro \& Benjamin 1991, and Schulman, Bregman, \& Roberts
1994) in contributing to our understanding of turbulent ISM heating
and the disk-halo connection.

At about the same time, observational discoveries on the nature of
several components of the ISM led to a improved understanding of its
overall phase structure (see the reviews in the book of van der Hulst
1997).  One important component is the warm neutral medium, WNM, or
the Lockman layer in our galaxy.  This component consists mostly of
small HI clouds, and diffuse (cirrus) material distributed in a
substantially thicker disk than the cold component (e.g., Dickey \&
Lockman 1990, Malhotra 1995, and Haynes \& Broeils 1997).  A second
component is the warm ionized medium (WIM), or diffuse ionized medium
(DIM), or the Reynolds layer in our galaxy, which is observed in
H$\alpha$ and other emission lines (e.g., Reynolds 1996, Rand 1997a).
These two media are continuous and overlapping, though the WIM extends
to greater scale heights above the disk than the WNM.

Both components have been studied in detail in our galaxy (see the
review of Dickey \& Lockman 1990, and the recent HI study of Malhotra
1995), and are probably common constituents of late-type galaxies (see
van der Hulst 1997).  The line emission of the WIM makes it the easier
component to study in other galaxies; it is observed in: 1) dwarf
irregular galaxies (e.g., Hunter \& Gallagher 1997), 2) edge-on disk
galaxies, especially the vigorously star-forming object NGC 891 (Howk
\& Savage 1997, Rand 1996, 1997a,b, 1998, Swaters, Sancisi \& van der
Hulst 1997), 3) large-scale outflows from nuclear starburst galaxies
(e.g., Lehnert \& Heckman 1996), and 4) nearby disk galaxies at
arbitrary inclinations (Wang, Heckman, \& Lehnert 1997).  These
studies also confirm the association of the various warm-to-hot
components (henceforth, collectively WHM) with star formation regions,
superbubbles, and large-scale outflows.  This association, in turn,
suggests connections with the still hotter ISM components observed in
X-rays and radio continuum.

The question of how the extensive WHM is heated (and how the WIM is
ionized) in galaxies that are not experiencing extensive starbursts is
not entirely answered.  Strong impulsive energy sources are quite
localized, while the WHM is not, in either our galaxy or others (Rand
1996, Wang, et al. 1997).  UV photofluxes can transfer energy over
long distances, and cosmic rays and magnetic fields can also
contribute to the pressure support.  However, a variety of evidence
indicates that other sources are needed to heat the WHM over most
scales.  The evidence includes the following: 1) fountain and
superbubble models, with a good deal of mechanical energy injection,
are very successful in accounting for the characteristics of the hot
halo gas (Shapiro \& Field 1976, Bregman 1980, Li \& Ikeuchi 1992,
McKee 1993), 2) an extra source of support beyond that associated with
random motions of typical clouds is needed to support the Galactic HI
(WNM) layer (Malhotra 1995), 3) the ``disturbed'' WIM in external
galaxies seems to require additional energy source (Rand 1997a,b, 1998,
Wang et al. 1997), 4) observations of Faraday rotation in our galaxy
(see summary in Minter \& Spangler 1996) demonstrate the existence of
turbulence on small scales in the diffuse ionized gas, and models
suggest that turbulent heating is important on those scales (Minter \&
Spangler 1997, Minter \& Balser 1997).  Turbulence, generated by the
impulsive sources and propagated by (magneto)acoustic waves and mass
flows may be the missing ingredient on the intermediate scales, as
well as the large and small scales.

The idea that interstellar cloud structure is turbulent is decades old
(see Larson 1979, Scalo 1987 and references therein).  Hydrodynamic
models of local regions of the ISM with heating and cooling sources
included clearly illustrate the development of turbulence (e.g., Rosen
\& Bregman 1995, Passot, Vazquez-Semadeni, \& Pouquet 1995).  Chappell
\& Scalo (1997) argue that clouds themselves are multifractal
manifestations of the interstellar turbulence. Elmegreen (1997) agrees
that interstellar clouds have a fractal structure, and further
proposes that the ``holes and gaps'' that make up the intercloud
medium are the result of turbulent heating rather than ``clearing'' by
supernova explosions.  Norman and Ferrara (1996) calculate that the
turbulent energy injection into the interstellar gas is characterized
by a very broad band spectrum, and that the general turbulent pressure
may exceed the thermal pressure by 1-2 orders of magnitude.  Thus,
there is increasing evidence that {\it turbulence supplies a large
part of the heating needed to maintain the continuous range of phases
in the ISM, and much of the pressure support.} This principle is the
central assumption on which the models described below are based.

\subsection{Large-Scale Structure}

Heating and cooling processes in the interstellar gas generally have
characteristic length scales of less than a kiloparsec or so.  These
local thermal processes co-exist with global regularities in the
structure of gas-rich disks, like the nearly universal surface density
profiles of the cold gas and stellar components.  The profile forms of
these components are often described as a negative exponential
functions of radius in the disk, though the gas surface density is
also well described as having a $1/r$ form (e.g., Struck-Marcell
1991).  Kennicutt's (1989, 1990, 1998a) influential studies of star
formation (henceforth SF) in galaxy disks showed that the neutral gas
surface density varies with radius in such a way that it is always
nearly equal to the radially dependent threshold density for
gravitational instability.  These results revived earlier suggestions
that local gravitational instabilities are needed to assemble the
massive clouds where star clusters are formed.  The threshold surface
density also varies as $1/r$ if the circular velocity is a constant,
independent of radius, as observed in many disks.

It is widely believed that the disk gas surface density profile is a
result of initial conditions, and the disk formation process.  The
reader is referred to the papers of Steinmetz \& M$\ddot{u}$ller
(1995), Dalcanton, Spergel, \& Summers (1997), and Mo, Mao, \& White
(1998) for recent discussions of disk formation and early evolution.
However, as noted above, thermal and turbulent pressure forces, as
well as gravity and centrifugal force, are important in the WHM, and
this makes it less likely that initial profiles are ``frozen out'' in
these media.  The effects of turbulent pressure on the large-scale
structure of the WHM, have not yet received much attention.  In the
context of the secular evolution of the cold gas, Struck-Marcell
(1991) pointed out that a $1/r$ surface density profile was required
to maintain the conditions of hydrostatic equilibrium in the disk with
minimal transport by shear viscosity (see below).  The same arguments
apply to the WHM, and more strongly, because of its shorter
sound-crossing timescale.  This leads us to the hypothesis that {\it
in any disk where the heating by SF activity is sufficient to maintain
and cycle a large fraction of the gas through the WHM, turbulent
hydrodynamic forces will regulate the gas surface density to the $1/r$
profile.}  Furthermore, the recent work of Martin (1998 and references
therein) suggests that substantial cycling rates are quite plausible.
This hypothesis provides a second pillar on which the models below are
built.

Dopita (1985) described one of the first models of global
self-regulation by star formation.  His model was based on the
assumption of equipartition between turbulent and thermal pressures.
In the models derived below we will consider cases where the turbulent
pressure exceeds the thermal pressure in the WHM phases, as suggested
by Norman and Ferrara.  This is analogous to the situation in the cold
cloud ensemble, where cloud random velocities are supersonic. We will,
however, assume pressure balance between different thermal phases.

Many mechanisms of self-regulation based on the intrinsic stability
properties of local star-cloud systems in disks have been suggested.
In particular, models with cloud buildup by collisional agglomeration
and disruption by SF activity have been popular, since they often
yield (Schmidt-type) SFRs with power-law dependences on local gas
density (e.g., Scalo \& Struck-Marcell 1984, Struck-Marcell \& Scalo
1987, Dopita 1990, Dopita \& Ryder 1994).  Paravano has extensively
investigated how star formation can be regulated by the thermal
conversion processes that operate between warm diffuse phases and the
small cool clouds (1988, 1989, also see Franco \& Shore 1984,
Diaz-Miller, Franco, \& Shore 1998, and Bertoldi \& McKee 1997 for
related models).  With reasonable approximations he finds that these
processes also produce a Schmidt Law SFR.

Silk (1997, and references therein) has argued for a somewhat
different star formation law, that includes a dependence on local
shear, and has described self-regulation and a derivation of the
Tully-Fisher relation in a model based on this law.  Kennicutt
(1998a,b) finds that the global properties of star formation in
galaxies are consistent with both the Schmidt and Silk/Wyse
phenomenological laws.

There are several difficulties with many of these approaches to
self-regulation.  The first is that most require some arbitrary
phenomenological assumptions, i.e., about SF or cloud collision rates,
equipartition, or constraints on the ambient pressure.  The second is
that most are local in the sense that they do not include the
hydrodynamic couplings to adjacent regions.  On the other hand, the
so-called chemodynamical models are based on a large-scale
hydrodynamical treatment.  For example, Samland, Hensler, \& Theis
(1997 and references therein) have recently presented two-dimensional
hydrodynamical models with three stellar components and two discrete
gaseous phases.  Inevitably, there are many uncertain parametrizations in
the couplings between the components, which limits the predictive
power of the models.  Moreover, they do not include the effects of
turbulent stresses in the intercloud medium, which we feel are
essential to an understanding of the large-scale structure.  The
primary role of these stresses distinguish the models below from
most previous ones.

\section{Global Analytic Models with Multiphase Turbulence}
\subsection{Densities and Radial Velocities} 

A good conceptual model of gas dynamics and star formation in galaxy
disks requires a quasi-static solution of multiphase hydrodynamic
equations that include the important thermohydrodynamic forces of
self-regulation.  Ideally, the model should be simple enough to allow
a clear understanding of both the structure of individual disks and
the universal relations between disks.  As a step towards this goal we
here introduce a model of a two-component ISM described by
cylindrically symmetric hydrodynamic equations, with turbulent stress
terms.  The two components are a cold isothermal (cloud) phase and a
mean WHM described by a locally adiabatic equation of state (see
below).  This two-phase model is a minimal description of the
multiphase interstellar gas, but it is sufficiently complex to capture
many interesting behaviors, as we will see below.  We will describe a
simple analytic version of the model here, but the model is readily
generalizable to a continuous range of phases, albeit with much
increased complexity in the equations.  We plan to develop that
generalization in a later paper.  In this paper we do not consider the
effects of magnetic forces or cosmic rays separately from the thermal
or turbulent pressures.

As part of the definition of a ``quasi-static'' disk, we will assume
that the mass exchanges between phases balance locally.  Then the
time-independent mass continuity equations are of the form,

\begin {equation}
{\Sigma_i}{v_{ri}}r = constant,
\end {equation}

\noindent where ${\Sigma_i}$ is the phase component surface density
and $v_{ri}$ is the component radial velocity, and with i = c,w for
the two phases.  For convenience, we will frequently use the
approximation that ${\Sigma_i} = {\rho_i}h_i$, where ${\rho_i}$ is the
component mass density, and $h_i$ is the component scale height in the
direction perpendicular to the disk.

We assume hydrostatic equilibrium in the vertical direction (i.e.,
perpendicular to the disk).  When the disk self-gravity dominates the
vertical potential gradient, then the solutions to the component
hydrostatic equations give component scale heights that increase
slowly with radius (as $r^{1/2}$, see Malhotra 1995 for a discussion of
the application of this approximation to the Milky Way).

The zero radial flow solution, $v_{rc} = v_{rw} = 0$, to equation (1)
is often assumed.  However, if ${\Sigma_i} {\propto} 1/r$, then there
is a more general family of solutions in which ${v_{ri}}(r) =
constant$.  Especially interesting are the solutions with $v_{rc} =
-({\Sigma_w}/{\Sigma_c})v_{rw}$, describing opposed radial flows in
the two components, but with zero net radial mass transport.  This
allows the slow inflow of high density cold (or cooling) gas to
replace gas heated into the warm phases by SF, which are on average
flowing outward.  The flow of individual mass elements might consist
of little more than a circulation within a local fountain (like a
transient convective cell), with the ensemble of fountains making up
the global inflow/outflow solution.  Figure 1 provides a schematic
view.

Because the scale-height ratio $h_c/h_w$ is small, inflow dominates in
the midplane of the disk, sandwiched between warm outflows above and
below.  The existence of two discrete radial velocities is a
consequence of the assumption of two distinct phases, and the model
can be generalized to a continuous range of phases as a function of
distance from the midplane with continuously stratified radial
velocities.  The fact that the $1/r$ profile is a good approximation
to the observed cold gas distributions in large parts of many
late-type galaxy disks (Struck-Marcell 1991) provides empirical
support for the constant radial velocity inflow$/$outflow solutions as
opposed to more complicated radius-dependent flows.  But why should
this profile be universal?

\subsection{Hydrodynamic Forces and the $1/r$ Surface Density Profile}

Several decades of numerical studies suggest that when the cold gas
density significantly exceeds the gravitational instability threshold
in all or part of a disk, then the result is the rapid development of
instabilities that generate readily observable clumps. (An example of
the clumping instability in disk formation is presented in Noguchi
1998.)  The massive clumps would generate strong SF and heating, which
would regulate the cold gas to lower densities by the processes
described above.  Thus, we do not expect to find densities much in
excess of the threshold, which scales as $1/r$ in a flat rotation
curve disk.

Hydrodynamical stability arguments also lead to a preference for the
$1/r$ surface density profile, as described by Struck-Marcell (1991).
For example, if the mean random velocity of the cold cloud ensemble is
a constant independent of radius (i.e., the ensemble is isothermal),
then only the $1/r$ profile yields a constant net pressure force
between adjacent annuli in the disk.  Any deviation from that profile
generates a pressure gradient proportional to the deviation, which
would allow the nonlinear amplification of disturbances.  This is not
consistent with a hydrodynamic steady state.  As noted above, this
argument is even stronger when applied to the WHM.

A qualitative, but very general argument, is based on the observation
that there is no obvious characteristic scale (other than the scale of
the rising rotation curve in the center) over the range extending from
a few kiloparsec to tens of kiloparsecs in late-type disks.  This
suggests that the important forces within the disk all have
essentially the same radial or distance scaling.  If not, we would
expect there to be observational signatures associated with changes in
the dominant force.  To make this point more definite, we write the
radial momentum equation for the cold gas component as follows,

\begin {equation}
{v_{rc}}\frac{\partial v_{rc}}{\partial r} = 
\biggl( \frac{-1}{\rho_c} \biggr)
\frac{\partial P_c}{\partial r} - 
\biggl( \frac{GM_o}{r_o} \biggr)
\frac{1}{r} + 
\frac{v_{{\phi}c}^2}{r},
\end {equation}

\noindent where $M_o$, $r_o$ are a gravitational scale mass and
radius, respectively, and $v_{{\phi}c}$ is the azimuthal velocity in
the cold component.  The terms on the right-hand-side represent the
pressure gradient, gravitational, the centrifugal accelerations.

In equation (2) the gravitational and centrifugal accelerations both
have a $1/r$ scaling.  There is no power-law solution for $v_{rc}$
that yields the desired scaling for the advection term.  However, when
$v_{rc}$ is constant this term vanishes.  Moreover, the isothermal
equation of state also yields the correct scaling in the pressure
gradient acceleration for any power-law density profile, though we
generally expect this term to be negligible in the cold gas.  Thus,
the adopted density and velocity profiles make up the only
self-similar family of steady solutions to equations (1) and (2).

The remaining momentum equations do not require any changes in the
density and velocity scalings. The azimuthal momentum equations are
described in the next section, the radial momentum equation for the
WHM is,

\begin {equation}
{v_{rw}}\frac{\partial v_{rw}}{\partial r} =
\biggl( \frac{-1}{\rho_w} \biggr)
\frac{\partial P_w}{\partial r} -
\biggl( \frac{GM_o}{r_o} \biggr)
\frac{1}{r} +
\frac{v_{{\phi}w}^2}{r} +
\alpha_w
\frac{{{\Delta}v_T}{{\Delta}v_{\phi}}}{r} ,
\end {equation}

This equation is very similar to the previous one, and in both cases
the left-hand-side is zero for the constant radial velocity
models. One difference is that we expect the pressure gradient term,
which contains both thermal and turbulent pressures, to be significant
in this component, in contrast to the cold component.  The additional
last term derives from an effective turbulent shear viscosity in the
azimuthal direction. This term must be included because the change in
specific angular momentum due to ``viscous'' forces changes the
effective centrifugal force from what we would expect in the absence
of ``viscosity''. Thus, in the warm gas, the non-negligible pressure
gradient, and the steady input of angular momentum allow gravity to be
balanced with less centrifugal force, while maintaining a constant
velocity outflow.

There is a corresponding rate of decrease of angular momentum in the
cool gas, but assuming that the total mass and angular momentum of the
cold gas are greater than the WHM, we have assumed that this term is
negligible in equation (2).  We will discuss the viscous shear terms
further in the next section, but we note here that the term in
equation (3) contains the quantity, ${\Delta}v_{\phi}$, which we
define as $max(v_{{\phi}c} - v_{{\phi}w}, {\Delta}v_T)$, and
${\Delta}v_T$ is the turbulent velocity dispersion in the WHM.

In this simple model we will assume that the viscous coefficients
${\alpha_i}$ in the two phases, the nearly circular azimuthal
velocities ${v_{{\phi}i}}$ and the velocity dispersions are all
constant.  In the case of the velocities this assumption is in accord
with observation, see van der Hulst (1997).  The pressure gradient is
due to the random motions of the cold gas elements, and not cloud
internal pressures.

\subsection{Viscosity and Pressure}

Many processes may contribute to the viscosity in the interstellar
gas, including: viscous shear within the cloud ensemble, drag against
the diffuse components, enhancements of these by bars and spiral
waves, turbulent and magnetohydrodynamic couplings, etc. At the same
time, turbulent pressure and turbulent angular momentum transfers to
the WHM can provide support against gravity, as well as driving the
radial flow.  As mentioned above, the support against gravity implies
that the circular velocity of this medium can be less than that of the
cold gas.  If the two (or multiple) phases have substantially
different rotation speeds, ${\Delta}v_{\phi}$, then we expect
turbulent shear viscosity between vertical layers to be the dominant
viscous term.  Cold gas heated by SF will be mixed into the WHM via
chimneys, fountains, and bubble shells, and will add angular momentum
to the WHM.  At the same time, cooling lumps or filaments of the WHM,
with lower specific angular momentum than the cold gas, will rain onto
the midplane, decreasing the angular momentum of the cool clouds that
sweep them up.  The net result will be somewhat like the friction
between two thin disks forced together at different speeds.

With the assumption that this is the only significant viscous term we
can write the steady azimuthal momentum equation as,

\begin {equation}
\frac{v_{ri}v_{{\phi}i}}{r} = 
-\alpha_i
\frac{{{\Delta}v_T}{{\Delta}v_{\phi}}}{r} ,
\end {equation}

\noindent where i=c,w as usual, and this equation is valid for both
components, though the radial velocities have different signs.  The
right hand term can be understood as dimensionally similar to a
Navier-Stokes kinematic viscosity term of the form -
${\nu}({{\partial^2}v_{\phi}}/{{\partial}z^2})$, with a viscosity
coefficient of order $\nu \simeq {\Delta}v_T{\lambda}$. In this case,
the ``mean free path'' $\lambda$ is of order the size of a typical
chimney, fountain, or the turbulent zone around an SF region, which we
assume to be roughly constant across the disk.  The vertical second
derivative we approximate as of order ${\Delta}v_{\phi}/{h_w^2}$,
where as noted above, the WHM scale height $h_w \propto r^{1/2}$. The
coefficient $\alpha_i$ is assumed to be of order unity. Although the
viscous term in the above equation may look somewhat unusual, in fact,
it is a variant of the usual $\alpha$-viscosity for supersonic
turbulence, $\nu \propto l_{turb}v_{turb}$, with scales appropriate to
this problem.

As noted above, we can understand the last term in equation (3) as a
momentum source term equivalent to a reduction in the centrifugal
acceleration due to azimuthal viscosity. We assume that the radial
velocities are small, and so, radial viscous accelerations are
negligible.  The centrifugal reduction can be treated as an
acceleration, which we estimate as,

\begin {equation}
\frac{{\partial}v_{ri}}{{\partial}t} 
\simeq 
\frac{v_{ri}v_{{\phi}i}}{r}
 = -\alpha_i
\frac{{{\Delta}v_T}{{\Delta}v_{\phi}}}{r} ,
\end {equation}

\noindent where the first equality is derived on the assumption that
this acceleration must have a magnitude sufficient to reduce the
azimuthal velocity to zero in a time $r/{v_{ri}}$.  The second
equality results from substituting the previous equation.  It is an
additional simplifying assumption that the coefficient on the
right-hand-side of this equation equals that of the previous equation,
though we expect them to be of the same order. This term is usually
neglected in viscous disk studies because $v_{ri}$ is small, but it is
important here. We will also see later that ${v_{rc}} << v_{rw}$, so
$\alpha_c << \alpha_w$, which justifies neglecting the term in
equation (2).

Because of the low densities and high temperatures that characterize
most of the WHM, we generally expect radiative cooling timescales to
be long, so in our simple model an adiabatic equation of state is more
appropriate than an isothermal one.  This raises a potential problem
with the scaling of the pressure gradient acceleration.  However,
shock heating plays an important role in determining the temperature
structure of the WHM, and the frequency and intensity of the shocks
varies with radius.  Since shocks generate entropy, this suggests that
the mean specific entropy of the WHM varies with radius, and so, the
adiabat of the warm gas must also vary with radius.  In this simple
model we write the equation of state for the warm gas as,

\begin {equation}
{P_{w,therm}}(r) = K(r) \rho_w^{\gamma}.
\end {equation}

\noindent where $P_{w,therm}$ is the thermal pressure, ${\gamma}
{\simeq} 5/3$, and the adiabatic constant K varies with radius.  (In
reality, it probably also varies with vertical height z.)  In
particular, we obtain the desired self-similar form for the pressure
gradient term with a radially dependent adiabatic constant of the
form, $K(r) {\propto} r$.  If the warm gas is flowing outward, this
increase in specific entropy would be the result of the cumulative
effects of shocks (see Appendix A).  It is interesting that {\it while
the warm phase is assumed to be locally adiabatic in this simple
model, the variation of pressure (and observables that depend on
thermal temperature) with radius is the same as a globally isothermal
gas.}  (Note that in terms of surface density and surface pressure we
have $\Pi = {K_S}\Sigma^{5/3}$, with $\Sigma \propto r^{-1}, \Pi
\propto r^{-1}$, and $K_S \propto r^{2/3}$, and again, $\Pi \propto
\Sigma$.)

At this point, the structure of a minimal model is nearly completely
defined by the empirically constrained ($1/r$) surface density
profiles, the adopted equations of state for thermal pressures, the
continuity equations, the radial and azimuthal momentum equations, and
the condition of pressure balance between phases.  Using the ideal gas
law to relate (total) pressure $P_w$ to the temperature $T_w$, we have
the following equation for the inter-phase pressure balance,

\begin {equation}
{\bf R}{\rho_w}{T_w}(1+\beta) = {\rho_c}({\Delta}v_c)^2 ,
\end {equation}

\noindent where the {\bf R} is the gas constant, ${\beta}(r)$ is a
correction factor for the turbulent contribution to the total pressure
in the WHM, and the right-hand-side represents the pressure in random
cold cloud motions (probably also turbulence dominated).  For
simplicity and consistency, we assume that $\beta$ is a constant, and
so, turbulent and thermal pressures scale with radius in the same way.

\section{Vertically-Dependent Circular Velocities and the Case of NGC
891} 

In this model the turbulent pressure gradient can provide some support
in the WHM against gravity, so the circular velocity of this medium
can be smaller than that of the cold gas.  Recently, Swaters, Sancisi,
\& van der Hulst (1997) discovered that the rotation velocities of HI
gas located in a plane parallel to, but above, the midplane of the
edge-on galaxy NGC 891 are less than those in the midplane by $25-100
km/s$. Over most of the disk the velocity difference was about $25
km/s$. The highest values come from a couple of points within a radius
of about 6.0 kpc. In the case of these points, the gas above the
midplane may have originated in an outflow from smaller radii, and
thus, considerably down the solid body part of the rotation curve.  If
so, the large velocity offset may be due to conservation of the small
angular momenta in this gas.

Swaters et al. provide several possible explanations for the general
effect, including local outflows and the angular momentum effect just
mentioned. They also mention the possibility of ``asymmetric drift'',
though they are skeptical that velocity dispersions as large as
suggested by the empirical formula for asymmetric drift in our galaxy
are achievable. However, since galactic asymmetric drift is a
phenomenon of collisionless stars on orbits with large epicycles, or
quite flattened elliptical orbits, we do not believe it is relevant
unless elements of the WHM can become isolated and collisionless for
large parts of their orbits.

Instead, we believe that the general velocity offset could be the
result of substantial turbulence in this vigorously star-forming
object (see Rand 1996, 1997a,b, 1998, and Howk \& Savage 1997).  We will
provide a simple numerical example below of how the model can account
for this phenomenon in this section. In fact, we view this phenomenon,
together with the vertical radial velocity gradient, as essential
predictions of the model.  Thus, detailed rotation studies, like that
of Swaters et al., in other nearby, edge-on disks with known ``DIG''
components like NGC 3079, NGC 4631, NGC 5775, NGC 4302 (Rand 1996 and
references therein), and NGC 55 (Ferguson, Wyse, \& Gallagher 1996,
Hoopes, Walterbos, Greenwalt 1996) should provide a strong test of the
model.

\section{Scaling the Model}

We can now describe how to scale the model, and provide some concrete
examples.  For brevity, we will not consider boundary conditions here,
but simply assume an infinite radius solution.  We begin by
initializing several observable quantities, and also the constant
radial mass transfer rate $( \dot M = 2{\pi}r{h_i}{\rho_i}v_{ri})$,
which is not generally observed.  However, this quantity is related to
the global SFR, since SF activity is responsible for the viscous and
pressure forces.  Numerical simulations should eventually yield a
relation between them.  Martin's (1998) studies of dwarf galaxies
found the intriguing result that ``shells lift gas out of the disk at
rates comparable to, or even greater than, the current galactic star
formation rates.''  This finding lends much credibility to the idea
that both radial mass transfer and phase exchanges occur in late-type
disks at interesting rates.

The remaining quantities we initialize are ${\rho_c}(r_o),
{\Delta}v_c(r_o), {\Delta}v_T(r_o)$, and ${\beta}$.  In principle,
these are all evaluated at a particular radius $r_o$.  In fact, the
last three are constrained to be constants, independent of radius.
The parameters determining the overall gravitational potential
(essentially $v_{{\phi}c}$) can be determined from observations of
stellar kinematics. Estimates for the values of ${\rho_c}(r_o)$ (or
${\Sigma_c}(r_o)$) and ${\Delta}v_c$ can be derived from HI
observations.  Like the radial mass flux, the values of ${\Delta}v_c$
for the cold clouds and ${\Delta}v_T$ of the WHM turbulence will be
determined by the SFR.  Thus, we expect that for a universal (e.g.,
halo) gravitational potential, {\it this family of model star-forming
disks is primarily two-dimensional, with the two dominant parameters
being the cool gas density at some point and the SFR.}

The values of the mean WHM turbulence parameters ${\Delta}v_T$ and
${\beta}$ are the most difficult to evaluate.  However, ${\beta}$
primarily enters the equations in the combination
$(1+{\beta})/{\beta}$, and so, whenever ${\beta} > a few$, turbulent
wave pressure dominates in the WHM and the exact value of ${\beta}$
does not greatly affect the other quantities.  Moreover, ${\beta}$ is
not independent of ${\Delta}v_T$ (both are functions of the SFR).  The
variable ${\Delta}v_T$, or a dimensionless combination like
${\Delta}v_T/v_{{\phi}c}$, can be viewed as a scaling parameter of the
model, like the Mach number in shock hydrodynamics.  Different mean
WHM values, appropriate for galaxies with different SFRs, are obtained
by changing the value of this parameter, as we will demonstrate.

Given our initial parameter values, we can now derive the scaling
equations for the remaining variables.  First, from the condition of
vertical hydrostatic balance, we get the cool gas scale height,

\begin {equation}
h_c = \Biggl[ \frac{{\Delta}v_c^2}{{\pi}G({\rho_*}+{\rho_c})}
\Biggr] ^{1/2},
\end {equation}
where $\rho_*$ is the local star density. (The stars are assumed to
have a larger scale height than the cool gas, so $\rho_*$ is
essentially constant.)

Next, we use the continuity equation to derive $v_{rc}$, the cool gas
inflow, and from equation (2) we derive the value of $v_{{\phi}c}$

\begin {equation}
v_{{\phi}c} = \sqrt{\frac{GM_o}{r_o}}.
\end {equation}

This determines the last of the cool gas parameters, and we proceed to
the remaining WHM quantities.  Using equation (7) and the definition
of $\beta$, we note that,

\begin {equation}
{\beta}{\bf R}{T_w} = {{\Delta}v_T}^2.
\end {equation}

\noindent This equation can be solved directly for $T_w$, but it also
allows us to solve for the ratio of component scale heights.  The
scale height ratio equals the square root of the ratio of effective
temperatures (see expression for $h_c$ above), which includes both a
thermal and turbulent part in the case of the WHM.  These parts are
related by the previous equation, and so, we can write,

\begin {equation}
\frac{h_w}{h_c} =
{\biggl(}\frac{1 + \beta}{\beta}{\biggr)}^{1/2}
\frac{{\Delta}v_T}{{\Delta}v_c}.
\end {equation}

\noindent Using the temperature expression above in the pressure
equation (eq. (7)), we get the related result for the component
density ratio,

\begin {equation}
\frac{\rho_c}{\rho_w} =
{\biggl(}\frac{1 + \beta}{\beta}{\biggr)}
\frac{{\Delta}{v_T}^2}{{\Delta}{v_c}^2}.
\end {equation}

\noindent The previous two equations can now be combined to give the
component surface density ratio,

\begin {equation}
\frac{v_{rw}}{-v_{rc}} = 
\frac{\Sigma_c}{\Sigma_w} =
{\biggl(}\frac{1 + \beta}{\beta}{\biggr)}^{1/2}
\frac{{\Delta}v_T}{{\Delta}v_c},
\end {equation}

Finally, we return to the radial momentum equation (eq. 3) to
determine $v_{{\phi}w}$. the left-hand side of this equation is zero,
assuming constant mean radial velocities, and for the pressure
gradient term the radial dependence of the pressure in this model ($P
\propto \rho \propto r^{-3/2}$) allows us to write,

\begin {equation}
\frac{r}{\rho_w}
\frac{{\partial}P_w}{{\partial}r} =
- \frac{3P_w}{2\rho_w} =
-\frac{3}{2} 
{\biggl(}\frac{1 + \beta}{\beta}{\biggr)}
{{\Delta}v_T}^2.
\end {equation} 

\noindent Then substituting from equation (4) for the $\alpha$ term,
multiplying by r, and dividing by ${v_{{\phi}c}}^2$, the radial
momentum equation can be written,

\begin {equation}
{\biggl(}\frac{v_{{\phi}w}}{v_{{\phi}c}}{\biggr)}^2 +
{\biggl(}\frac{v_{rw}}{v_{{\phi}c}}{\biggr)}
{\biggl(}\frac{v_{{\phi}w}}{v_{{\phi}c}}{\biggr)} -
{\Biggl[} 1 - \frac{3}{2}
{\biggl(}\frac{1 + \beta}{\beta}{\biggr)}
{\biggl(}\frac{{\Delta}v_T^2}{v_{{\phi}c}^2}{\biggr)}
{\Biggr]} = 0.
\end {equation}

\noindent where we have also used the approximation that
$GM_o/({r_o}{v_{{\phi}c}}^2) = 1.$ This is a Bondi-Parker,
accretion/wind equation, and is quadratic in the variable
$(v_{{\phi}w}/v_{{\phi}c})$.  It is a central result of the model.

To evaluate that equation numerically, we use equation (13) for
$v_{rw}$, and the continuity equation, with a given mass flux,
for $v_{rc}$. As an example, let us assign the following
representative values:

\noindent $M_o = 2 \times 10^{11} M_{\odot},$
\ $r_o = 10 kpc.,$ 
\ ${\rho}_c(r_o) = 3.0 amu/cm^3,$
\ ${\Delta}v_c = 6.0 km/s,$ 
\ ${\Delta}v_T = 30 km/s,$ 

\noindent $\beta = 3,$ 
\ a mass flow of $\dot M = 2.0 M_{\odot} yr.^{-1},$ and
a stellar density of ${\rho_*} = {\rho}_c$
 
\noindent (used for computing scale heights).  

Then we derive values of 

\noindent $h_c = 130 pc.,$ 
\ $h_w = 770 pc.,$
\ ${\rho_w}/{\rho_c} =0.030,$ 
\ $v_{rc} = -3.4 km/s,$ 
\ $v_{rw} = 20 km/s,$

\noindent ${v_{{\phi}_c}} = 290 km/s,$ 
\ ${v_{{\phi}_w}} = 277 km/s,$
 and $T_w = 21,000K.$  

Note that the radial velocities are very small compared to the
azimuthal velocities, and also less than the turbulent velocity
dispersions.  Thus, we expect radial velocities to be quite difficult
to observe.

If instead of the above value for ${\Delta}v_T$, we substitute a
higher value of ${\Delta}v_T = 50 km/s$, we obtain values of $h_w =
1300 pc., v_{rw} = 33 km/s$, and ${\Delta}v_{\phi} = 25 km/s$.  The
thicker WHM layer and the larger azimuthal velocity difference in this
case are both in accord with the observations of NGC 891.  The surface
density of the WHM relative to that of the cool gas increases
linearly with ${\Delta}v_T$, consistent with the idea that the strong
WHM emission in NGC 891 is the result of strong turbulence driven by
the vigorous SF.

On the other hand, we noted above that Swaters et al. (1997) did not
see such high velocity dispersions in their HI observations of NGC
891. However, these authors note that in the more nearly face-on
galaxy NGC 6946, vertical velocities of up to $100 km/s$ were detected
by Kamphuis \& Sancisi (1993).  These, and related observations, can
be readily understood if the turbulent motions of HI gas in the thick
disk and halo are primarily vertical. This situation is very natural
if the most of the high dispersion HI gas is either entrained in local
fountains and outflows, or is in the form of ``high velocity clouds''
consisting of cooled halo material falling back onto the disk (e.g.,
Benjamin 1999, Benjamin \& Danly 1997).

The impressive recent study of Thilker, Braun \& Walterbos (1998) on
large HI shells in NGC 2403 also provides input on this
question. These authors find mean in-plane shell expansion velocities
of $26 km/s$, and individual cases extending up to $56 km/s$.  We
expect that much of the turbulent energy has already been vented in
these large bubbles, and that the shells are observed in a
deceleration stage.

Recent optical observations also provide evidence for the vigorous
turbulence required by our model. Wang, Heckman \& Lehnert's (1997)
spectroscopic study of the DIM in half a dozen nearby disks led them
to suggest the existence of two components. The first is a 'quiescent
DIM' with low ionization states and line widths of $20-50 km/s$,
versus the high ionization state, 'disturbed DIM' with line widths of
$70-150 km/s$. They suggest that the former is photoionized by diffuse
O star radiation, while the latter is mechanically heated by
supernovae and winds.  In sum, turbulence with values of ${\Delta}v_T
\simeq 50 km/s$ or greater seems in accord with recent observations,
and seems to yield very reasonable model values for actively
star-forming disks. 

\section{Star Formation Properties and Other Regularities in the
Family of Models}

In constructing the model above, we have implicitly assumed that the
SF law above threshold is constrained by a self-regulated heating and
cooling balance.  In the simplest case, we assume that all the
important heating processes are directly proportional to the SFR
(e.g., O star photoheating and turbulent wave heating).  The important
cooling processes in the WHM include: 1) adiabatic cooling of high
pressure gas elements, 2) radiative cooling in the mean WHM with a
rate proportional to the WHM density squared, and 3) turbulent shock
dissipation in the WHM, which depends on the SFR and gas density
squared.  The last two cooling rates generally scale with mean WHM gas
density squared, and thus, require a similar density dependence in the
SFR. That is, a Schmidt Law on average, albeit in density, rather than
surface density.  However, the processes involved are sufficiently
complex that deviations from an m=2 density power would not be
surprising.

Thus, we write the following schematic equation for the heating and
cooling balance for regions above the SF threshold,

\begin {equation}
{f_{SF}}{\rho_c^2} = {n_w^2}\sigma{c^3}.
\end {equation}

\noindent The left-hand-side of this equation represents the Schmidt
Law heating.  The right-hand-side is a schematic collisional
dissipation term, with ``cross section'' $\sigma$ and ``sound speed''
c. For example, this term could represent radiative cooling in the WHM
via collisional excitation, with the temperature dependence contained
within the factor ${\sigma}{c^3}$.  However, if we make the
approximation that $c \propto {\Delta}v_T$, and assume that $f_{SF}$
is universal, then the equation provides a scaling for the net
dissipation cross section $\sigma$ in terms of ${\Delta}v_T$ and
${\rho_c}/{\rho_w}$. (See Appendix A for further discussion of the
constrained SFR.)

There is a substantial literature on individual interstellar heating
and cooling terms on many scales (e.g., see the recent discussions of
Norman \& Ferrara 1996, and Ferriere 1998).  Nonetheless, we still
have a long way to go to fully understand the broad-band heating and
cooling terms in the ISM.  Any more specific formulation of the
balance equation would probably require the introduction of insecure
parametrizations with.  It would also require additional physics,
beyond that included in the simple model considered here.  We will not
pursue these topics in this paper, but merely point out that they may
be easier to study in the context of the well-defined global structure
provided by the model.

Equations (8)-(15) show that the model is a simple similarity
solution to the hydrodynamic equations.  In the limit of small radial
velocities and small turbulence (and thus little WHM component) the
model must be essentially the same as the self-similar, viscous,
Mestel disks studied recently by Bertin (1997) and Mineshige \&
Umemura (1996). (Our model also has some similarities to the
one-component convective model of Waxman (1978).)  The self-similar
structure of the model helps to understand the universal properties of
gas-rich galaxy disks, like the Tully-Fisher relation between maximum
circular velocities and the total luminosities of disk galaxies (e.g.,
Courteau 1997 and references therein), and universal gas density
profiles (see e.g., the extensive HI study of Broeils \& Rhee 1997).

Eisenstein and Loeb (1996) point out that the small dispersion in the
observational Tully-Fisher relation suggests the operation of a
``strong feedback process'' that ``regularize(s) SF and gas
dynamics,'' like the model presented here.  On the other hand, Mo et
al. (1998) believe that the small Tully-Fisher scatter could in fact
come out of early galaxy formation processes.  Even so, Eisenstein and
Loeb are probably also correct if a large fraction of galaxies
experience merger events subsequent to their formation. That is, while
we expect galaxy collisions and mergers to disrupt the ``universal''
disk structure, turbulent self-regulation will re-establish it.

Some of these questions are answered by the new numerical
hydrodynamical models of galaxy formation in several cosmologies by
Elizondo et al. 1999). These models included multiple gas phases and
supernova feedback. The authors found that the Tully-Fisher scatter of
their model galaxies was within the acceptable both with and without
the feedback effects included. However, only the feedback models
reproduced the correct slope of the Tully-Fisher relation, and the
slope was quite sensitive to the feedback amplitude. 

Another regularity, Freeman's Law, states that high surface brightness
galaxies all have about the same central surface brightness, or that
disk galaxies have a maximum central surface brightness (see Courteau
1996 and references therein). This too would seem to be a natural
consequence of large-scale SF regulation, albeit in the central
regions of the disk.  In a number of nearby starburst galaxies the gas
surface density continues to follow a power-law as far into the center
as it can be resolved (Struck-Marcell 1991, Young et al. 1995).  In
many cases this may be the result of gas inflow driven by a bar
component or other disturbance.  In many other late-type disks the
gradient in the surface density flattens to a value of about 10 solar
masses per square parsec in the central, rising rotation curve region
(though often with a central, molecular gas spike, see Young et
al. 1995).  We speculate that this latter case represents a normal
quiescent state.  Disks of both profile types are observed to have a
comparable value of $\Sigma_c$ at the radius where the rotation curve
flattens (Broeils \& Rhee 1997).  We conjecture that Freeman's Law may
be the result of the fact that most stars form in the centers of
bright, late-type disks at a rate appropriate to this gas density, and
over comparable timescales.

\section{Conclusions}

The following list summarizes the properties and some probable
consequences of the models presented in this paper.

1) A $1/r$ surface density profile is assumed in both thermal phases
(on the basis of the Least Dissipation Principle and other arguments,
see section 2.2). It is also assumed that vertical scale heights are
determined by local self-gravity, and so, increase slowly with radius
(as $\propto r^{1/2}$). This implies that mean volume densities scale
as, $\rho_i \propto r^{-3/2}$.

2) Thus, the ratio of gas phase densities are constant across the
star-forming region of the galaxy disk. The model predicts that the
value of this ratio depends on the amount of turbulence, and
specifically, on the parameter ${\Delta}v_T/{\Delta}v_c$.

3) We assume that the circular velocities of each gas phase are
constants independent of radius.  The equations of state and the
assumption that all momentum equation terms have the same radial
scaling implies that the remaining velocities, $v_{rc}, v_{rw},
{\Delta}v_T, {\Delta}v_c$, are also constant with radius.

4) In general, the model allows all of the velocities in each phase -
azimuthal, radial, and dispersive - to have different (non-zero)
values.  The NGC 891 effect of different rotational velocities as a
function of height above the disk (Swaters et al. 1997) is predicted
to be generic in turbulent disks.

5) The model predicts a hierarchy in velocity magnitudes in each
phase, i.e., azimuthal velocities $>>$ velocity dispersions $>$ radial
velocities.  The low value of the latter will make them difficult to
observe. This hierarchy is the result of the similarity equations
and the simple wind/accretion equation (eqs. (8) - (15)).

6) However, even such low velocity radial flows are consistent with
mass fluxes comparable to typical SFRs in late-type disks.  In the
absence of radial flows, we would expect gas consumption at smaller
disk radii, to modify the gas density profile. Radial replenishment
can prevent this and effectively distribute the consumption across the
disk.  The radial flow may also draw on reservoirs of gas in the
non-star-forming outer disk, further increasing the global consumption
time.  Evolutionary effects will be considered in a later paper.

7) Similarly, galactic abundance gradients will be smoothed by the
large-scale radial flows.  The simple closed-box model of chemical
evolution within isolated disk annuli is not appropriate in the
context of these radial circulation models.  However, the quantitative
effects of radial flow are complicated by the fact that the flows are
slow.  E.g., with a radial flow velocity of order $3 km/s$, the
timescale for a gas element to cross a disk of radius $10 kpc$ is a
few billion years.  Moreover, motion of a gas element will generally
be partly advective and partly diffusive in this turbulent
environment. Thus, the typical smoothing time may be only a little
less than the typical disk age.

8) The model requires a balance between heating and cooling.  Heating
is primarily the result of SF activity, and most cooling terms depend
on the second power of the gas density (assuming the constant phase
balances of the model).  Thus, a Schmidt Law SFR is the natural result
of the thermohydrodynamical balance.

9) If the Schmidt Law (or a related parametrization, such as that of
Silk and Wyse, see Silk 1997) is in fact a consequence of global
hydrodynamic self-regulation, then there are some immediate
corollaries.  Perhaps, the most important is that transient, burst
modes of SF are possible when disturbances take galaxies far from the
regulated state.  Thus, SF phenomenologies may be very different in
highly disturbed galaxies (e.g., Struck-Marcell \& Scalo 1987), or
during galaxy formation.  On the other hand, the Schmidt
parametrization may be marginally valid in environments where,
$r/{\Delta}v_T = {\tau_{relax}} < {\tau}$, with $\tau$ defined as an
appropriate ``age''.  For example, the Schmidt Law may work in waves
in both grand design and collisional galaxies if the wave crossing
time is longer than $\tau_{relax}$.  The same argument may be valid in
the centers of major merger remnants.

10) Global regularities in star-forming disks, such as the
Tully-Fisher relations and the Freeman Law, may also be the result of
global self-regulation, of the type inherent in the present
models. They may also be the result of formation processes, including
turbulent self-regulation during formation.  Continuing
self-regulation is important for maintaining the global regularities,
and restoring them following a disturbance.

11) The model equations can readily be generalized beyond the
two-phase version described here to include a continuous range of
phases.  More sophisticated treatments of viscosity, turbulence,
heating and cooling processes can be included, much as detailed
nuclear rates and opacities are included in stellar evolution models.

In sum, the hydrodynamic similarity model presented above is an
attempt to bring together the essential thermohydrodynamical processes
needed for a coherent conceptual picture of actively star-forming
galaxy disks as self-regulated, multiphase, ``dissipative
structures.''  The basic hydrodynamic structure of each phase is much
like that of an isothermal polytrope, but these are not quasi-static,
equilibrium states.  The model assumes that there are turbulent flows
on many scales, and the WHM is more accurately viewed as a set of
locally adiabatic states, with specific entropy gradients.  The disk
structure described by this model, with gas profiles regulated in
accord with minimal dissipation and transport, and SF simultaneously
sustained and moderated by slow, radial flows, probably describes most
late-type disk galaxies.

Acknowledgements: We are very grateful to Julia K. Burzon for creating
Figure 1.  We also want to thank P. N. Appleton, R. Benjamin,
V. Charmandaris, B. Elmegreen, L. Sage, J. Scalo, J. M. van der Hulst
and especially an anonymous referee and our editor S. Shore for
providing helpful (regulating!) feedback. We are also grateful for
partial page charge support from an American Astronomical Society
Small Grant.

\appendix {Appendix A:
A Plausibility Argument for Radial Entropy Increase and
the Schmidt Law}

The positive radial entropy gradients in the WHM component of the
model presented above may seem unphysical, especially since this gas
is expanding in an average outward flow.  However, there is an obvious entropy
source in the nonlinear acoustic waves that partially support this
flow. At the same time, radiative cooling provides an obvious entropy
sink, yet when this gas experiences significant cooling it is
generally transformed into the cool component. Thus, on average, the
entropy of gas that stays in the WHM either increases, or is balanced
by adiabatic expansion.

Specifically, consider a non-cooling element of the WHM moving outward
in the mean flow above the midplane of the disk.  Suppose for
simplicity, that its specific entropy is significantly increased only
when it passes directly over a young star cluster, assuming such star
clusters are the primary source of shock turbulence. Then, the rate at
which the entropy of that element is increased will be proportional to
the number of young clusters it passes over per unit time. (Note that
because of the reduced azimuthal velocity of the WHM, the element will
pursue a spiral trajectory as viewed in a reference frame comoving
with the midplane gas.)

If the SFR is described by a Schmidt Law, and assuming most new stars
are born in clusters, then the number of clusters within a thin
annulus of width ${\Delta}r$, at radius r is,

\begin {equation}
N_{cl} = 2{\pi}r{\Delta}r
\biggl( \frac{\psi}{M_{cl}} \biggr)
{\tau_{cl}} = {c_1} \frac{{\Delta}r}{r^{1+m}}.
\end {equation}

In this equation, $\psi$ is the usual SFR (mass of stars produced per
unit area per unit time), $M_{cl}$ is the mean cluster mass, and
$\tau_{cl}$ is the mean lifetime of the massive stars in the
cluster. The final equality assumes a Schmidt Law of the form, $\psi
\propto {\Sigma_{c}^{2+m}}$, and a surface density profile of the form
$\Sigma_c \propto 1/r$.  The constant $c_1$ is the combination of all
constants in the previous equality.

We further assume that the gas element expands in the azimuthal
direction by an amount proportional to r as it moves outward. This is
just the expansion that is required to maintain the assumed surface
density profile. It also guarantees that the WHM element covers a
constant fraction of each thin annulus it crosses.  Thus, the gas
element crosses a constant fraction of the clusters $N_{cl}$ in each
thin annulus, and it is reasonable to assume that the rate of shock
hits and entropy increase it experiences is proportional to this
annular cluster fraction. Thus, the net entropy increase in traveling
from radius $r_1$ to radius r is

\begin {equation}
{\Delta}S \simeq {c_2} {\int} 
\frac{dr}{r} \simeq {c_2} log(r/{r_1}),
\end {equation}
in the case m = 0.  The constant $c_2$ contains the product of the
earlier constant $c_1$, the annulus fraction covered by the gas
element, and the mean entropy input per cluster.

For a perfect gas,

\begin {equation}
S = log{\bigl(} P/{\rho^{\gamma}} \bigr) 
= log({K_V}(r)),
\end {equation}
where the second equality makes use of equation (6), in section 2.3,
(and the subscript V emphasizes that these are volume quantities).
Assuming that this equals the preceding equation to within an additive
constant, we have,

\begin {equation}
{K_V}(r) \propto {r^{c_2}}.
\end {equation}

As noted in section 2.3 the quasi-steady model presented above
requires that ${c_2} = 1$.  One factor contained within the constant
$c_2$ is the magnitude of the SFR, e.g., the SFR at a particular
radius.  Thus, the self-regulating feedback processes can adjust the SFR
in such a way that ${c_2}$ is driven towards unity, giving the desired
entropy gradient.

Therefore, the fact that the entropy gradient is a power-law follows from
the $m=0$ spatial dependence of the Schmidt Law SFR, while the value
of the power depends on the magnitude of the SFR.  In other words, the
entropy gradient required for a hydrodynamic steady state can be
achieved by feedback adjustments to the amplitude and spatial
dependence of the SFR.

Now let us consider the effects of these self-regulating processes
from a slightly different point of view.  According to equations (6)
and (7) the mean WHM temperature scales as, $T_w \propto (1 +
\beta){K_V}{\rho^{2/3}}$. For a globally adiabatic gas, K is constant,
and $T_w \propto 1/r$ (for constant $\beta$).  While, as described
above, for a locally adiabatic gas, with a radial entropy gradient
such that $K_V \propto r$, we have $T_w$ constant. If the latter
alternative does not obtain, then the variation of scale height with
radius will be different than the $h \propto r^{1/2}$ form assumed
above. Qualitatively, if the scale height increases less rapidly, then
warm gas remains closer to the midplane and is denser, so cooling
rates are increased. If more of this gas goes into the cool phase, we
expect that the SFR will increase (relative to the quasi-steady
model), driving increased turbulence and heating, and increasing the
scale height.

Conversely, if the scale height increases more rapidly with radius,
the SFR will be less than in the steady model, eventually diminishing
pressure support, and reducing the scale height.  More generally,
because of the temperature dependence of the cooling rates,
significant temperature gradients would likely result in pressure
imbalances, which would lead to time-dependent convection (as well as
thermal conduction). That is, temperature gradient states do not
generally satisfy the steady state equations above.

These considerations are qualitative, and have loopholes, but they
do show why global states with modest entropy gradients would be
preferred. They also provide some insight into how closely the Schmidt
Law SFR is connected to such states.

FIGURE CAPTIONS

Fig. 1. - Schematic illustrating local fountain flows, cooling
filaments, and the global radial flows in a model disk (courtesy Julia
K. Burzon)


\begin{references}

\reference{benj1999} Benjamin, R. A. 1999, in Stromlo Workshop on
High-Velocity Clouds, ASP Conference Series 166, eds. B. K. Gibson \&
M. E. Putnam (ASP: San Francisco), p. 147

\reference{benj1997} Benjamin, R. A. \& Danly, L. 1997, \apj, 481, 764

\reference{bert1997} Bertin, G. 1997, \apj, 478, L71

\reference{breg1980} Bregman, J. N. 1980, \apj, 236, 577

\reference{brin1990} Brinks, E. 1990, in The Interstellar Medium in
Galaxies, eds. H. A. Thronson, Jr. \& J. M. Shull (Dordrecht: Kluwer),
p. 39

\reference{broe1997} Broeils, A. H., \& Rhee, M.-H. 1997, \aap, 324,
877 

\reference{bert1997} Bertoldi, F., \& McKee, C. F. 1997, Rev. Mex. AAp
Serie de Conferencias, Vol. 6, 1st Guillermo Haro Conference on
Astrophysics: Starburst Activity in Galaxies, p. 195

\reference{chap1997} Chappell, D. W., \& Scalo, J. M. 1997, \apj, 
submitted (preprint: astroph 9707102)

\reference{cour1996} Courteau, S. 1996, \apjs, 103, 363

\reference{cour1997} Courteau, S. 1997, \aj, 114, 2402

\reference{cox1981} Cox, D. P. 1981, \apj, 245, 534

\reference{dalc1997} Dalcanton, J. J., Spergel, D. N., \& Summers,
F. J. 1997, \apj, 482, 659

\reference{diaz1998} Diaz-Miller, R. I., Franco, J., \& Shore,
S. N. 1998, \apj, 501, 192

\reference{dick1990} Dickey, J. M. \& Lockman, F. J. 1990, \araa, 28,
215 

\reference{dopi1985} Dopita, M. A. 1985, \apj, 295, L5

\reference{dopi1990} Dopita, M. A. 1990, in The Interstellar
Medium in Galaxies, eds. H. A. Thronson, Jr. \& J. M. Shull
(Dordrecht: Kluwer), p. 437

\reference{dopi1994} Dopita, M. A., \& Ryder, S. D. 1994, \apj, 430,
163 

\reference{eise1996} Eisenstein, D. J., \& Loeb, A. 1996, \apj, 459,
432 

\reference{eliz1999} Elizondo, D., Yepes, G., Kates, R., Muller, V.,
\& Klypin, A. 1999, \apj, 515, 525

\reference{1997} Elmegreen, B. G. 1997, \apj, 477, 196

\reference{ferg1996} Ferguson, A. M. N., Wyse, R. F. G., \& Gallagher,
J. S., Jr. 1996, \aj, 112, 2567

\reference{ferr1998} Ferriere, K. 1998, \apj, 497, 759.

\reference{fran1984} Franco, J., \& Shore, S. N. 1984, \apj, 285, 813

\reference{hayn1997} Haynes, M. P. \& Broeils, A. H. 1997, in The
Interstellar Medium in Galaxies, ed. J. M. van der Hulst (Dordrecht:
Kluwer), p. 75

\reference{heil1984} Heiles, C. 1984, \apjs, 55, 585

\reference{hoop1996} Hoopes, C. G., Walterbos, R. A. M., Greenwalt,
B. E. 1996, \aj, 112, 1429

\reference{howk1997} Howk, J. C., \& Savage, B. D. 1997, \aj, 114,
2463 

\reference{hunt1997} Hunter, D. A., \& Gallagher, J. S. III 1997,
\apj, 475, 65

\reference{kamp1993} Kamphuis, J. J., \& Sancisi, R. 1993, \aa, 273,
L31 

\reference{kenn1989} Kennicutt, R. C., Jr.  1989, \apj, 344, 685

\reference{kenn1990}  Kennicutt, R. C., Jr. 1990, in The Interstellar
Medium in Galaxies, eds. H. A. Thronson, Jr. \& J. M. Shull
(Dordrecht: Kluwer), p. 405

\reference{kenn1998a} Kennicutt, R. C., Jr.  1998a, \apj, 498, 541

\reference{kenn1998b} Kennicutt, R. C., Jr.  1998b, in Galaxies:
Interactions and Induced Star Formation, Saas Fee Advanced Course 26,
eds. D. Friedli, D. Martinet, \& D. Pfenniger (Berlin: Springer), p. 1 

\reference{lars1979} Larson, R. B. 1979, \mnras, 186, 479

\reference{lehn1996} Lehnert, M. D., \& Heckman, T. M. 1996, \apj,
462, 651

\reference{liik1992} Li, F., \& Ikeuchi, S. 1992 \apj, 390, 405

\reference{macl1989} Mac Low, M., McCray, R., \& Norman, M. L. 1989,
\apj, 337, 141

\reference{malh1995} Malhotra, S. 1995, \apj, 448, 138

\reference{mart1998} Martin, C. L. 1998, \apj, 506, 222

\reference{mcke1993} McKee, C. F. 1993, in Back to the Galaxy,
eds. S. S. Holt \& F. Verter (New York: A. I. P.), p. 499

\reference{mcke1977} McKee, C. F., \& Ostriker, J. P. 1977, \apj, 218, 148

\reference{mine1996} Mineshige, S., \& Umemura, M. 1996, \apj, 469,
L49 

\reference{mint1996} Minter, A. H., \& Spangler, S. R. 1996, \apj,
458, 194

\reference{mint1997a} Minter, A. H., \& Balser, D. S. 1997, \apj, 484,
L133 

\reference{mint1997b} Minter, A. H., \& Spangler, S. R. 1997, \apj,
485, 182

\reference{moma1998} Mo, H. J., Mao, S. \& White, S. D. M. 1998,
\mnras, 295, 319

\reference{nogu1998} Noguchi, M. 1998, \nat, 392, 253

\reference{norm1996} Norman, C. A., \& Ferrara, A. 1996, \apj, 467,
280 

\reference{norm1989} Norman, C. A., \& Ikeuchi, S. 1989, \apj, 345,
372 

\reference{parr1988} Parravano, A. 1988, \aap, 205, 71

\reference{parr1989} Parravano, A. 1989, \apj, 347, 812

\reference{pass1995} Passot, T., V\'azquez-Semadeni, E., \& Pouquet,
A. 1995, \apj, 455, 536

\reference{rand1996} Rand, R. J. 1996, \apj, 462, 712

\reference{rand1997a} Rand, R. J. 1997a, in The Interstellar Medium in
Galaxies, ed. J. M. van der Hulst (Dordrecht: Kluwer), p. 105

\reference{rand1997b} Rand, R. J. 1997b, \apj, 474, 129

\reference{rand1998} Rand, R. J. 1998, \apj, 501, 137

\reference{reyn1996} Reynolds, R. J. 1996, in The Physics of Galactic
Halos, eds. H. Lesch, R.-J. Dettmar, U. Mebold, \& R. Schlickeiser
(Berlin: Akademie Verlag), p. 57

\reference{rose1995} Rosen, A., \& Bregman, J. N. 1995, \apj, 440, 634 

\reference{saml1997} Samland, M., Hensler, G., \& Theis, Ch. 1997,
\apj, 476, 544

\reference{scal1987} Scalo, J. M. 1987, in Interstellar Processes:
Proceedings of the Symposium, eds. D. J. Hollenbach \& H. A. Thronson,
Jr. (Dordrecht: Reidel), p. 349

\reference{scal1984} Scalo, J. M., \& Struck-Marcell, C. 1984, \apj,
276, 60.

\reference{schu1994} Schulman, E., Bregman, J. N., \& Roberts,
M. S. 1994, \apj, 423, 180

\reference{shap1991} Shapiro, P. R., \& Benjamin, R. A. 1991, \pasp,
103, 923

\reference{shap1976} Shapiro, P. R., \& Field, G. B. 1976, \apj, 205,
762 

\reference{silk1997} Silk, J. 1997, \apj, 481, 703

\reference{stei1995} Steinmetz, M., \& M$\ddot{u}$ller, E. 1995,
\mnras, 276, 549

\reference{stru1991} Struck-Marcell, C. 1991, \apj, 368, 348

\reference{stru1987} Struck-Marcell, C., \& Scalo, J. M. 1987, \apjs,
64, 39

\reference{swat1997} Swaters, R. A., Sancisi, R., \& van der Hulst,
J. M. 1997, \apj, 491, 140

\reference{teno1990} Tenorio-Tagle, G., Rozyczka, M., \& Bodenheimer,
P. 1990, \aap, 237, 207

\reference{thil1998} Thilker, D. A., Braun, R., \& Walterbos,
R. A. M. 1998, \aa, 332, 429.

\reference{tomi1992} Tomisaka, K. 1992, \pasj, 44, 177

\reference{vand1996} van der Hulst, J. M. 1996, in The Minnesota
Lectures on Extragalactic Neutral Hydrogen, A.S.P. Conf. Series 106,
ed. E. D. Skillman (San Francisco: A. S. P.), p. 47

\reference{vand1997} van der Hulst, J. M. (ed.) 1997, The Interstellar
Medium in Galaxies (Dordrecht: Kluwer)

\reference{wang1997} Wang, J., Heckman, T. M., \& Lehnert, M. D. 1997,
\apj, 491, 114

\reference{waxm1978} Waxman, A. M. 1978, \apj, 222, 61 

\reference{youn1995} Young, J., et al. 1995, \apjs, 98, 219

\end{references}
\end{document}